\documentclass[11pt]{article}

\usepackage[normalem]{ulem}
\usepackage{amsmath, amssymb,amsthm,bm,bbm,fullpage}
\usepackage{mathrsfs}
\usepackage{colortbl}
\usepackage[all]{xy}
\usepackage{epsfig}
\usepackage{subfigure}
\usepackage{graphics}
\usepackage{multirow}
\usepackage{lineno}
\usepackage{setspace}
\usepackage{caption}

\usepackage{natbib}
\usepackage{graphicx}

\theoremstyle{remark} 
	
	\doublespace

\begin{document}

\begin{centering}
{\huge
\textbf{The natural selection of good science}
}
\bigskip
\\
Alexander J. Stewart$^{1,*}$ and Joshua B. Plotkin$^{2,*}$
\\
\bigskip
\end{centering}
\begin{flushleft}
{\footnotesize
$^1$ Department of Biology, University of Houston, Houston, TX, USA
\\
$^2$ Department of Biology, University of Pennsylvania, Philadelphia, PA, USA
\\
$^*$ E-mail: astewar6@central.uh.edu; jplotkin@sas.upenn.edu
}
\end{flushleft}
\noindent\textbf{Scientists in some fields are concerned that many, or even most, published results are false. A high rate of false positives might arise accidentally, from shoddy research practices. Or it might be the inevitable result of institutional incentives that reward publication irrespective of veracity. Recent models 
and discussion of 
scientific culture predict selection for false-positive publications, as research labs that publish more positive findings 
out-compete more diligent labs. There is widespread debate about how scientific practices should be modified to avoid this degeneration. Some analyses suggest that ``bad science" will persist even when labs are incentivized to undertake replication studies, and penalized for publications that later fail to replicate. Here we develop 
a framework for modelling the cultural evolution of research practices that allows labs to expend effort on theory -- enabling them, at a cost, to focus on hypotheses that are more likely to be true on theoretical grounds. Theory restores the evolution of high effort in laboratory practice, and it suppresses false-positive publications to a technical minimum, even in the absence of replication.
In fact, the mere ability choose between two sets of hypotheses -- one with greater chance of being correct than the other -- promotes better science than can be achieved by having effortless access to the better set of hypotheses. 
Combining theory and replication can have a synergistic effect in promoting good scientific methodology and reducing the rate of false positive publications. Based on our analysis we propose four simple rules to promote good science in the face of pressure to publish.}


\section*{Introduction}

Scientists are concerned about the state of science. There is ample evidence to suggest that in some fields a large portion of reported results 
may be false \citep{kerr1998harking,ioannidis2005most,simmons2011false,john2012measuring,simonsohn2014p,rahal2015estimating,begley2015reproducibility}. The quality and magnitude of empirical evidence for this concern varies across disciplines and is a matter of debate \citep{munafo2017manifesto}. But there is widespread acceptance that ``researcher degrees of freedom" -- such as flexibility in study design, measurement, and reporting -- can lead to a high rate of false-positive reports. 
The dominant view holds that, in some fields, 
a sizable portion of published findings are false -- a viewpoint publicized so widely that the lay person may reasonably be suspicious of the scientific enterprise. To remedy this situation, there have been remarkable efforts to fund and undertake large-scale replication studies to help identify errors in the literature and understand how they arise from current scientific practice \citep{rahal2015estimating,klein2018many,ebersole2016many,camerer2018evaluating}.
Among other approaches, such as pre-registration and registered reports \citep{nosek2015promoting,begley2015reproducibility,munafo2017manifesto,nosek2018preregistration,munafo2018}, a balance between testing new hypotheses and replicating published studies is now proscribed as a matter of course \citep{munafo2017manifesto}.

At the same time, models for the cultural evolution of scientific practice have suggested that replication efforts will not suffice to arrest the inevitable trend towards increasing false-positive rates driven by incentives to publish positive results \citep{smaldino2016natural,Grimes:2018aa}. Several authors have instead called for increased attention to theory as key to restoring a healthy scientific practice \citep{smaldino2016natural,muthukrishna2019problem, smaldino2019better}. Whether, or how, theory will actually promote good science  -- that is, reduce the rate of false-positive reports -- has not been studied in a formal framework. Moreover, models of cultural evolution used to interrogate the value of replication have been investigated primarily by simulation, without systematic mathematical analysis. 
Here we work to address both of these outstanding issues in the meta-scientific literature.

There are two ways that theory can aid scientific inquiry. When a field of research is underpinned by a well developed body of theory, the community of scientists can focus on those hypotheses that are more important or have a greater prior chance of being correct. That is, theory can give all researchers easier access to stronger hypotheses. At the same time, even in fields where a theoretical framework is not yet well developed or widely accepted, an individual lab that expends effort to model the system they are researching will generate stronger hypotheses by clarifying and quantifying their intuitions and by weeding out unlikely or illogical hypotheses. We show that this latter process -- individual labs expending effort to select stronger hypotheses -- has profound consequences for the cultural evolution of scientific practice.

Our analysis generalizes earlier models for the evolution of scientific practice in response to  pressure to publish positive results. In particular, we extend earlier work by analyzing the possibility that individual labs may expend effort on ``theory'' to improve the quality of the hypotheses they choose to test. We analyze our model both mathematically and by simulation, showing that the pressure to publish does \textit{not} produce an inevitable decline in the quality of science provided effort can be expended on theory. Rather, the system becomes bi-stable: it can support either high-quality science (low rates of false-positive reports) aided by theory, or a decline towards low-quality science and minimal effort. We quantify the basins of attraction towards these two different outcomes. Then we show how interventions such as replication can facilitate the stability of good-science equilibria. Finally we offer four simple rules, arising from our analysis, to promote good science in the face of pressures to publish.

\section*{Model}

Methods from cultural evolution can be applied to study the development of research practices in response to institutional incentives \citep{mcelreath2015replication,smaldino2016natural,Grimes:2018aa}. This approach rests on the idea that competing research groups vary in methodological traits that affect their success and that are ``heritable" either by differential imitation \citep{Traulsen:2006zr} or by differential production of students who then form their own labs, adopting the practices of their mentors. 
\\
\\
\noindent\textbf{Efficacy and Effort:} In order to study the natural selection of good science we adapt the model of \cite{smaldino2016natural}, which characterizes a research lab in terms of its ``power'' and ``effort''. Power and effort are treated as traits that can evolve in the population of labs via a process of natural selection. Together these traits determine the rate at which a given lab generates novel results for publication, which is a natural proxy for success (i.e.~fitness) in the face of inter-lab competition and the pressure to publish positive results.

Power in this context refers to the overall efficacy of the techniques a lab uses to identify true associations, and thus it describes the entire process of testing a hypothesis and producing a publication. \cite{smaldino2016natural} note that increasing a lab's power also increases its rate of false positives, unless effort is exerted; and that increasing effort decreases the productivity of a lab, because it takes longer to perform rigorous research. 

In order to avoid confusion with the familiar concept of statistical power, we distinguish the efficacy of a lab's techniques, denoted $V$, from the actual statistical power with which the lab correctly identifies true hypotheses, denoted $P(+|T)$. The statistical power of a lab is a function of \textit{both} the efficacy $V$ and the effort $e$ the lab puts into research (Figure 1). Following \cite{smaldino2016natural} we assume $V\in[0,1]$ and $e\in[1,\infty)$.

Under this model the process of producing a research paper proceeds as follows: (i) A lab selects a hypothesis to test. (ii) If the hypothesis is in fact true, the lab identifies it as such with a probability $P(+|T)$, which depends on the efficacy of the lab's techniques and the effort exerted to test the hypothesis. However, if the hypothesis is in fact false the lab mis-identifies it as true with a probability $P(+|F)$, which again depends on the efficacy of the lab's techniques and on the effort they exert. (iii) If the hypothesis is labelled as true the work is published  -- that is, we assume that only positive results are published \citep{smaldino2016natural}.

We assume that both true and false positive rates increase with a lab's efficacy, $V$, and decrease with the lab's effort, $e$. Effort decreases the rate of positive results because greater effort is tantamount to a more conservative and rigorous approach to hypothesis testing. We choose the following functional forms for the rates of true and false positives in terms of effort $e$ and efficacy $V$,
\begin{align}
\nonumber &P(+|T)=\frac{V}{\gamma}\times\frac{\gamma e}{1+\gamma(e-1)}\\
&P(+|F)=\frac{V}{\theta}\times\frac{1+(\theta-1)e}{1+(\theta-V)(e-1)}.
\end{align}
\\
According to this formulation the true positive rate increases linearly with efficacy, but it declines with effort \textit{i.e.}~as the lab becomes more conservative. The false positive rate is a convex increasing function of efficacy \citep{smaldino2019better}, as more powerful techniques increase the chance of false positives; but this can be counterbalanced by increasing effort. Just as in \cite{smaldino2016natural}, increasing efficacy always increases publication rate, namely the rate of positive findings.

Our formulation for the rate of true and false positives generalizes the model of 
\cite{smaldino2016natural}. The two formulations are identical in the limit $\theta=\gamma=1$. In general, however, our formulation differs from \cite{smaldino2016natural} in an important way: effort expended to reduce false positives also has the effect of reducing true positives (for $\gamma>1$), whereas \cite{smaldino2016natural} assumed the true positive rate is independent of effort. This more general formulation avoids a pathology that was present in earlier work: the tautological limit of $P(+|T)\to1$ and $P(+|F)\to1$ occurs only when efficacy is maximized ($V\to1$) \textit{and} effort is minimized ($e\to1$) under our model. This tautological limit corresponds to a situation where a lab simply labels all hypotheses as true, and so it should occur only when a lab expends minimal effort.

Our formulation also generalizes \cite{smaldino2016natural} in the limit of maximum effort, $e\to\infty$. This limit produces $P(+|T)\to1/\gamma$ and $P(+|F)\to1/\theta$, where the parameters $1/\theta$ and $1/\gamma$ define the technical limits of true and false positive discovery in a given field of research. 
\\
\\
\noindent\textbf{Hypothesis Selection:} The rate at which a lab discovers positive results depends on the true and false positive rates (Eq.~1) as well as the underlying probability that a hypothesis the lab selects to test is true, $P(T)$. One way to imagine science is as a ``grab bag" of hypotheses, each of which is true with a fixed probability $b$. We might imagine scientists as reaching into the bag, eyes closed, and drawing a hypothesis which they then test in the manner described above \citep{smaldino2016natural}. 

For many labs though, hypothesis selection is itself a product of effort. This effort may consist of broad engagement with the prior literature, which highlights some hypotheses as more plausible than others based on consistency with established results across many fields of science. 
Alternatively, it may consist of a lab can expending effort to produce models and theory, which enable the production of systematic and self-consistent predictions that can be tested as empirical hypotheses. 

In order to describe the process of putting effort into hypothesis selection we assume
\begin{equation}
P(T)=\frac{b_0+b_1(e-1)}{e}
\end{equation}
where $b_0$ is the baseline rate of true hypotheses under the ``grab bag'' model. According to this formulation, effort $e\geq1$ expended on hypothesis selection increases the prior probability that a selected hypothesis is true from the baseline rate $b_0$ to a maximum value $b_1>b_0$, achieved when a lab puts maximum effort into the development of theory and engagement with prior literature (Figure 1).
\\
\begin{figure}[th!] \centering \includegraphics[scale=0.3]{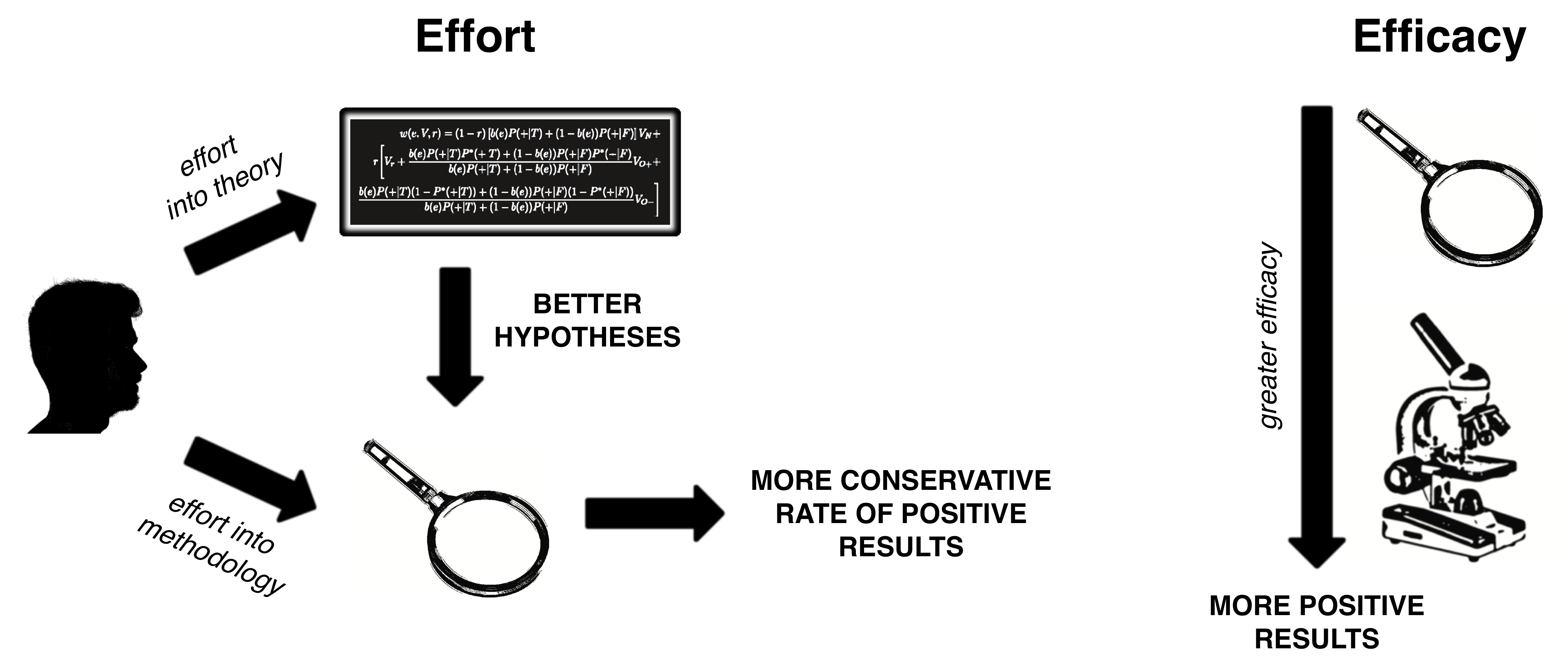}
\caption{\small \textbf{How can a lab do better science?} Science can be made better in two basic ways: 1) A lab can expend more effort, which means (all other things equal) that the lab selects a hypothesis with a higher prior probability of being correct and that, at the same time, the lab is more conservative about testing the hypothesis.
2) A lab can develop more effective methods, which means (all other things equal) that the rate of positive results increases.}
\end{figure}

\noindent\textbf{Publication and Replication:} To study the impact of replication on the cultural evolution of scientific practice, we assume that each lab chooses to replicate a published study at rate $r$, rather than attempting to produce a novel study \citep{smaldino2016natural}. The outcome of each replication attempt depends on the standing body of published literature and it is complicated to describe analytically. However, replication can be analyzed concisely under the simplifying assumption that labs all experience replication of their work at the same rate. We analyze this case mathematically, and we later show via simulation that our analytic results are good approximations even when this assumption is relaxed.

A lab produces novel results at a rate $\rho$,
\begin{equation}
\rho=(1-\eta\log_{10}(e))\times(1-r)\times(P(T)P(+|T)+P(F)P(+|F)),
\end{equation}
where the term $(1-\eta\log_{10}(e))$  describes the time it takes to produce a piece of research using effort level $e$ \citep{smaldino2016natural}. Similarly labs engage in replication studies at rate
\begin{equation}
\phi=(1-\eta\log_{10}(e))\times r
\end{equation}
where we assume that all replications are publishable regardless of outcome \citep{smaldino2016natural}.
%
\\
\noindent\textbf{Adaptive Dynamics of Science:} We can analyze the natural selection of good science via the payoffs associated with publication of novel results and replication of previous results. We first analyse the evolution of scientific practice under the simplifying assumptions of adaptive dynamics. In this framework an infinite population of labs are assumed to use identical strategies, and the success of a new strategy $i$, which differs slightly from the norm, is tested against the current resident strategy \citep{Mullon:2016aa,Leimar}. 
The expected fitness of a lab with a novel strategy $i$, denoted $w(e_i,V_i,r_i)$, is approximated by (see SI):
\begin{align}
 w(e_i,V_i,r_i)=\rho_iB_N+\phi_iB_r+
\frac{1}{2}\frac{\rho_i\phi}{l}(p_{i}B_{O+}-q_{i}C_{O-})
\end{align}
where $l$ is the ratio of published material in the corpus of the field to the number of active labs
(see SI). $B_N$ is the payoff for publishing a novel result, $B_r$ the payoff for publishing a replication study, $B_{O+}$ the payoff for having another lab successfully reproduce your work, and $C_{O-}$ the cost of having another lab fail to reproduce your work.  Here, $p_i$ and $q_i$ give the probability that a replication attempt by another lab on a study produced by lab $i$ is successful or unsuccessful respectively (see SI section 1.1).


The equilibria of the adaptive dynamical system occur when the selection gradient is zero

\begin{align}
\nonumber &\frac{\partial w}{\partial e_i}\Big|_{e_i=e}=0\\
&\frac{\partial w}{\partial r_i}\Big|_{r_i=r}=0
\end{align}
\\
Note that $\frac{\partial w}{\partial e_i}\Big|_{V_i=V}>0$ for all $V$, and so the system necessarily evolves to maximum methodological efficacy, $V=1$, as in the model of \cite{smaldino2016natural}. Eq.~6 cannot in general be solved analytically (see SI section 1), but it can be systematically explored numerically. 
\\
\\
\noindent\textbf{Individual-based Simulations:} In addition to mathematical analysis by adaptive dynamics, we also perform Monte Carlo simulations in polymorphic, finite populations, where lab strategies replicate according to a copying process \citep{Traulsen:2006zr}. We assume that science is produced according to Eqs.~1-4 
and that replication can occur once for any study present in the corpus, which has size $L$. Labs are assumed to become inactive when they adopt a new strategy, which may be thought of as retirement of a senior professor and replacement by a new hire. When a new lab is formed we assume that mutations perturb effort $e$, efficacy $V$, and replication rate $r$. Mutational perturbations are drawn uniformly from $[-0.01,0.01]$, and mutations occur at rate $\mu_e$, $\mu_V$ and $\mu_r$ respectively (see SI for full details).

First we perform simulations in the limit $\gamma=\theta=1$, where our model coincides with \cite{smaldino2016natural}. In this limit we can reproduce the findings of \cite{smaldino2019better}: bad science evolves in the absence of theory (SI Figure S5). Next we analyze the ability of theory, replication and methodological efficacy to preserve good science. 

\subsection*{Results}

\noindent\textbf{Theory produces good science:} 
When a lab cannot improve hypothesis selection by effort, then science will evolve to a state where labs simply label all novel hypotheses as true -- that is, the evolution of bad science \citep{smaldino2016natural}. As we show below, however, the mere act of expending effort to find stronger hypotheses is sufficient to stabilize good science. We define good science as an equilibrium that maintains a false positive rate close to the technical minimum, $P(+|F)\sim V/\theta$. Under our model this can occur only when effort is high (Eq.~1).

We model the act of expending effort to find stronger hypotheses via Eq.~2, where minimum effort ($e=1$) results in selection of a hypothesis with prior probability $P(T)=b_0$ and maximum effort ($e\to\infty$) results in a hypothesis with prior probability  $P(T)=b_1>b_0$. That is to say, a lab can expend effort to identify stronger hypotheses -- meaning those that have a higher probability of being true.

Expending effort to find stronger hypotheses produces good science, whereas simply having effortless access to stronger hypotheses does not (Figure 2). The figure shows the results of simulations in three different regimes: (i) only weak hypotheses available ($b_0=b_1=0.01$) (ii) only strong hypotheses available ($b_0=b_1=0.25$) and (iii) choice, via effort, between weak and strong hypotheses ($b_0=0.01$ and $b_1=0.25$). In the first two cases bad science evolves, with effort declining to its minimum and true and false positive rates increasing to unity, which replicates the simulation results of \cite{smaldino2016natural}. 
However in the third case, when effort can be expended to select stronger hypotheses, we find something quite different. As labs evolve, effort \textit{increases} from its initial value to a level that maintains a high true-positive rate and a low false-positive rate -- the evolution of good science. Note that expending effort on theory in order to select strong hypotheses produces a good-science equilibrium even when effortless access to equally strong hypotheses leads to bad science.

\begin{figure}[th!] \centering \includegraphics[scale=0.2]{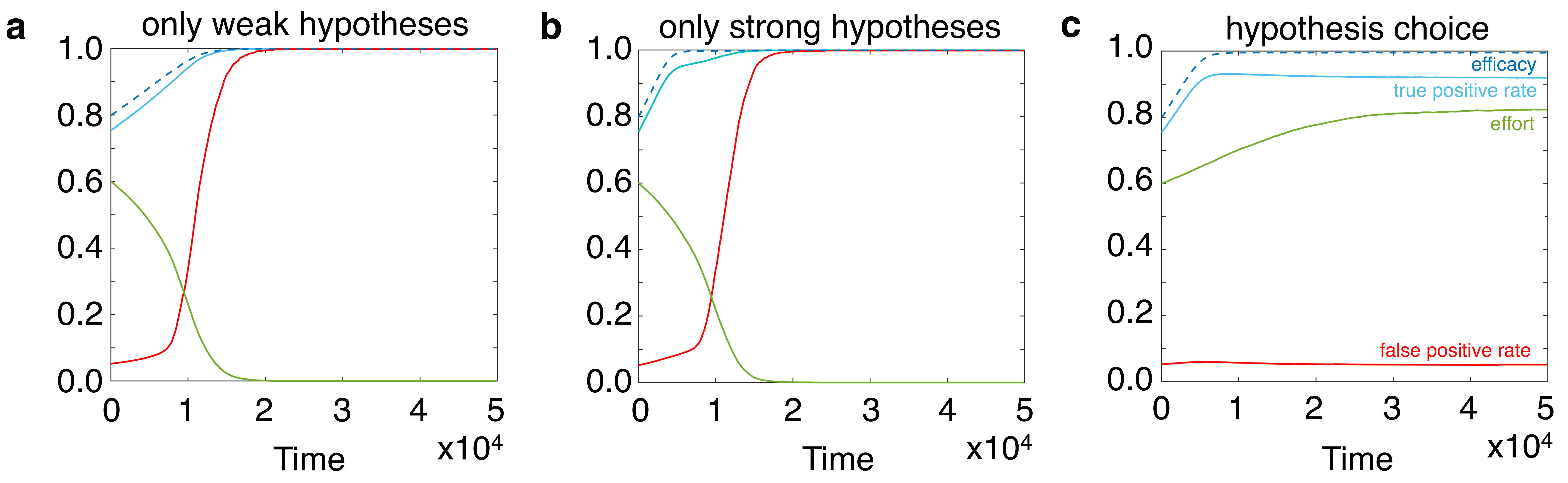}
\caption{\small \textbf{The evolution of good science:} We ran individual-based simulations in which $N=100$ labs compete to publish positive results, in the absence of replication. In each panel we plot the trajectories of efficacy $V$ (dashed blue line), true positive rate $P(+|T)$ (solid blue line), false-positive rate $P(+|F)$ (red line), and effort $(e-1)/e$ (green line). a) When only weak hypotheses are available ($b_0=b_1=0.01$) efficacy increases over time, but effort declines, so that the population evolves to a bad-science equilibrium. b) The same is true when only strong hypotheses are available ($b_0=b_1=0.25$). c) When effort can be put into choosing between weak and strong hypotheses ($b_0=0.01$ and $b_1=0.25$) a stable, good-science equilibrium emerges, and effort and efficacy both increase, leaving the false positive rate close to the technical minimum $P(+|F)\sim 1/\theta$. The figures show the mean trajectories over an ensemble of $10^3$ replicate simulations. The rate of publication for each lab was determined by Eqs.~1-3; mutations occurred to effort $e$ and efficacy $V$ at rate $\mu_e=\mu_V=0.01$. Mutational perturbations to efficacy were drawn uniformly from the range $[-0.01, 0.01]$, and effort was assumed to change by $\pm 1$ upon mutation. Cultural evolution occurred via a copying process (see SI), payoffs were set at $B_N=1$, and $\theta=25$.}
\end{figure}

How does expending effort on hypothesis selection promote good science? Analysis of our model by adaptive dynamics (Eqs.~5-6 and SI section 1) shows that when effort can be expended to find stronger hypotheses, the system becomes bi-stable (Figure 3). The bad-science equilibrium identified by \cite{smaldino2016natural} always exists, but another equilibrium emerges that features high effort and low false positives. For a broad range of parameters the basin of attraction towards the good-science equilibrium is much larger than the basin of attraction towards the bad-science equilibrium. 

The reason why increasing effort can be advantageous once theory is introduced, is that greater effort results in a greater probability of testing a true hypothesis in the first place, $P(T)$. Near to the good-science equilibrium, decreasing effort tends to reduce the overall rate of publication, because it makes hypotheses less likely to be true a priori; and the lab still puts effort into assessing the veracity of each hypothesis, so that they end up identifying more hypotheses as false, thereby reducing publication rate. This phenomenon is sufficient to stabilize good science. 

\begin{figure}[th!] \centering \includegraphics[scale=0.175]{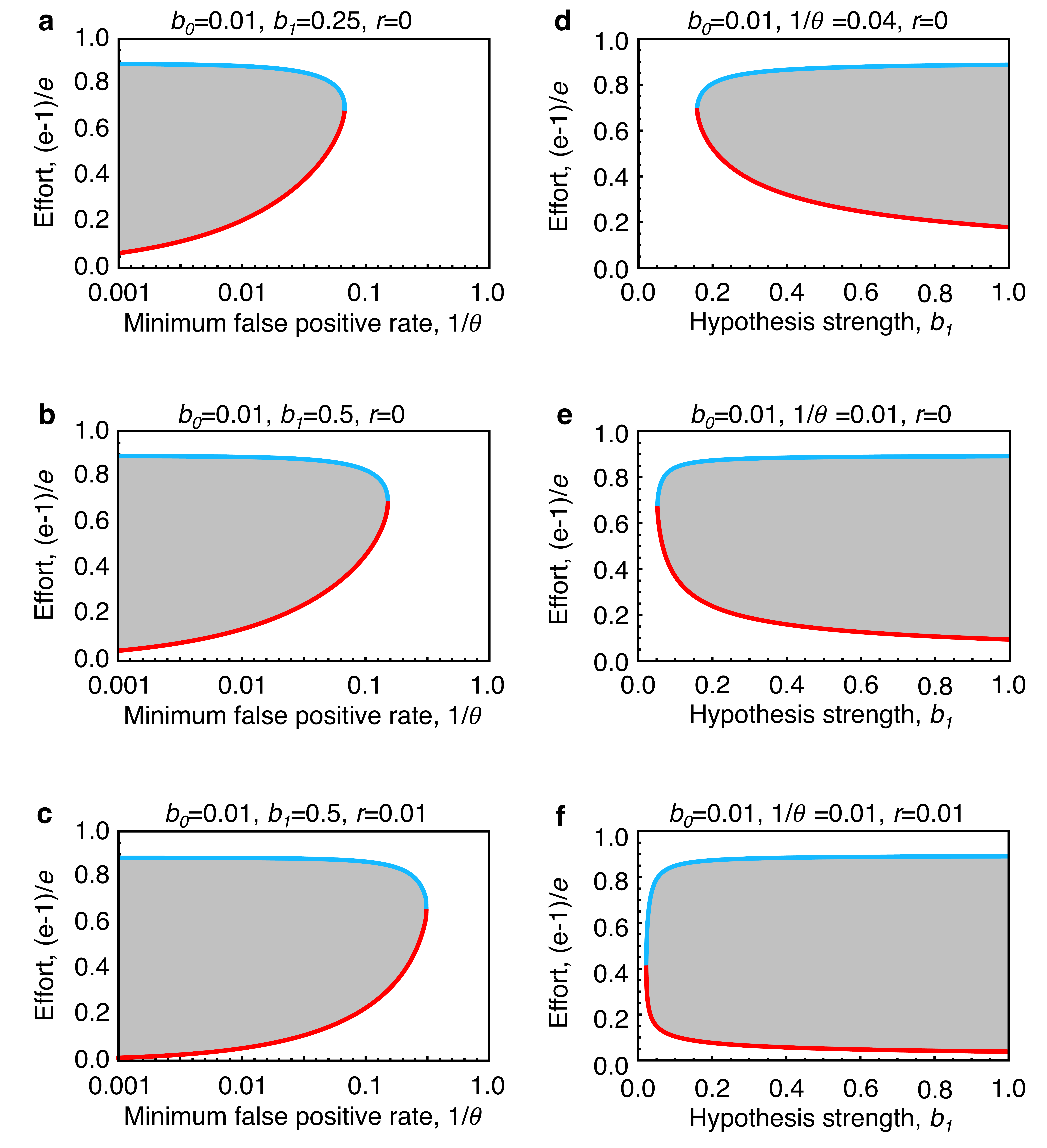}
\caption{\small \textbf{Analysis of equillibria by adaptive dynamics.} The figure shows equilibrium publication strategies in a large population of labs, as a function of model parameters. Plotted in each panel are the locations of the stable (blue) and unstable (red) equilibria as a function of either the technical minimum false positive rate $1/\theta$ (left column) or the maximum achievable hypothesis strength $b_1$ (right column). For many parameter choices the system is bi-stable, with a good-science equilibrium indicated by the blue line and a bad-science equilibrium at minimum effort $(e-1)/e=0$. In the gray regions selection favors increasing effort towards the good-science equilibrium; whereas in the white regions selection favors ever decreasing effort towards to bad-science equilibrium. a) For $b_0=0.01$ and $b_1=0.25$ and without replication ($r=0$), stable good science requires a technical minimum true positive rate no greater than $1/\theta=0.08$. b) With better theory, meaning the possibility of stronger hypotheses $b_1=0.5$,  good science is stable  with even lower methodological efficacy (e.g.~$1/\theta>0.1$). c) Adding replication at a low rate ($r=0.01$) enables good science to be maintained for even larger values of $1/\theta$. Similar patterns occur when we fix $1/\theta$ and vary $b_1$ (right column): increasing methodological efficacy allows good science to emerge even with weaker hypotheses (panels d-e), and replication decreases the need for strong theory even further (panel f). Payoffs are set at $B_N=1$, $B_r=0.2$, $B_{O+}=0.1$ and $C_{O-}=100$ and $l=5$.}
\end{figure}

The emergence of good science as a stable response to the pressure to publish depends on the extent to which a field has developed along three major axes (Figure 3 and Figure S3). (i) A field must have achieved a sufficient degree of methodological efficacy ($1/\theta$ sufficiently small). That is, if low rates of false positives cannot possibly be achieved even through high effort, then good science cannot be maintained. (ii) Labs must have sufficient ability to discriminate between strong and weak hypotheses ($b_1$ sufficiently larger than $b_0$). That is, good science cannot be maintained if a field does not yet have sufficient theory to allow labs the ability to select stronger hypotheses through effort. (iii) Good science can be stabilized when labs undertake replication ($r>0$), which can help make up for weaker methodology ($\theta$) or theory ($b_1$), but this is not always guaranteed (see SI)

\clearpage

\noindent\textbf{Replication can make good science easier:} Theoretical and methodological development are intrinsic characteristics of a field. While both tend to improve over time, they cannot be varied exogenously. In contrast, the rate of replication can be increased or decreased exogenously by introducing institutional incentives or policies that require replication \citep{munafo2017manifesto}. And so much of the debate over how to promote good science has been focused on encouraging replication and similar interventions \citep{munafo2017manifesto,munafo2018}. 

\begin{figure}[th!] \centering \includegraphics[scale=0.2]{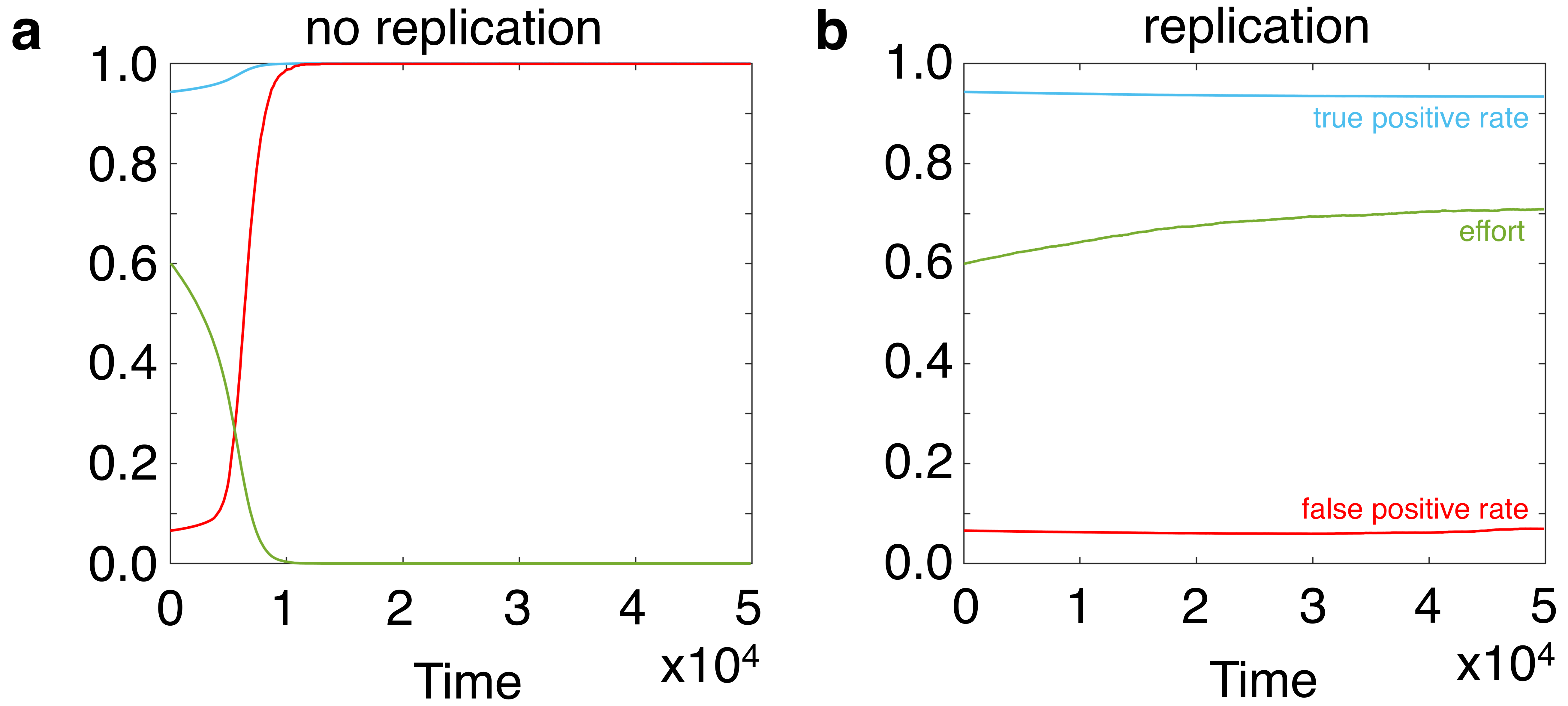}
\caption{\small \textbf{Synergy between replication and theory.} The figure shows results of individual-based simulations for the evolution of scientific practice with and without replication. In the regime $1/\theta=0.04$, shown here, theory and replication are both required to produce good science, as predicted by mathematical analysis by adaptive dynamics (Figure 3a). (a) In the absence of replication, both true (blue) and false (red) positive rates increase to unity, and effort declines to a minimum $(e-1)/e=0$ (green). b) However, when replication occurs at a rate $r=0.1$, effort increases over time towards a good-science equilibrium in which false positives are rare. All parameters are the same as in Figure 2c, except for $\theta$. Replications are chosen from a corpus of $L=10^5$ novel studies, and each study is allowed to be replicated only once (see SI). Payoffs are $B_N=1$, $B_r=0.2$, $B_{O+}=0.1$ and $C_{O-}=100$.}
\end{figure}

Replication can help weed out bad science by re-testing published results and flagging the false positives. By imposing a cost on labs who publish false positives, replication reduces the incentive for labs to lazily label novel results as true without expending the effort to properly verify them. However, previous models for the evolution of scientific practice have found that replication cannot prevent the natural selection of bad science \citep{smaldino2016natural}. We too find that, in the absence of theory to enable hypothesis selection, replication alone does not produce good science (Figure 4). However we also find (Figures 3-4) that, in the presence of theory, replication can both increase the basin of attraction of good science and interact synergistically with stronger hypotheses and better methodology to stabilize good science. Figure 4 shows examples where the introduction of replication can make the difference between the evolution of bad versus good science.

We can also analyze the co-evolution of effort and replication rate. Using the framework of adaptive dynamics (Eqs.~5-6) we show that, when the cost for studies that fail to replicate is large ($C_{O-}\gg1$) both good- and bad-science equilibria persist, but replication is always lost (see SI Section 1.2 and Figure S1). Individual-based simulations produce similar results: in combination with theory replication evolves to low rates and good science is maintained, whereas without theory, replication cannot help to prevent the natural selection of bad science (SI Section 2.4 and Figure S4). Moreover, when replication occurs at a fixed rate and does not evolve, but is rather held in place as a policy, it can dramatically expand the basin of attraction of good science (SI Section 1.4 and Figure S3).


\section*{Discussion}

Scientific practice is amenable to scientific study. We have developed models of cultural evolution \citep{smaldino2016natural} to study how theory influences research effort and methodological efficacy for labs under pressure to publish. The ability to expend theoretical effort on hypothesis selection produces bi-stable dynamics: evolution will lead either to high-effort labs that publish reports with few false positives (good science), or alternatively to minimal-effort labs that publish results replete with false positives (bad science). Our mathematical analysis delineates which of these equilibria will arise, in terms of the payoffs for publication, the field's technical limits on true- and false-positive discovery, the payoffs associated with replication efforts, and the extent to which theory can improve hypothesis selection in the field.
\\
\\
\noindent\textbf{Four rules for the natural selection of good science:}
Our analysis suggests four simple rules to promote the evolution of good science, which offer both optimism and caution for researchers concerned about the publication of false results.  

\begin{enumerate}
    \item \textbf{Develop a robust theoretical framework.} A theoretical framework that enables labs to distinguish strong hypotheses from weak ones, at a cost, is sufficient to preserve good science. Science benefits from such a theoretical framework, in part \textit{because} it takes effort to identify stronger hypotheses.
    \item \textbf{Don't rely on replication alone.} Replication alone does not produce good science, but it can interact synergistically with theory to stabilize good science.
    \item \textbf{Different fields have different needs.} There is a trade-off between the methodological efficacy, theoretical sophistication, and the rate of replication required to sustain good science. A field that is more developed in one area can afford to be less developed in another, meaning better methods can make up for mediocre theory, to a point \citep{smaldino2019better}.
    \item \textbf{Bad science is always a danger.} Low-effort, attention-grabbing publication of any and all hypotheses is likely to be a stable equilibrium in all fields.
\end{enumerate}

\noindent\textbf{Models of models:} In our model for the evolution of scientific practice, increased effort makes almost everything harder: research takes longer and a lab is more conservative when labelling a hypothesis as true, both of which reduce the overall rate of publication. The only direct benefit of effort lies in selecting stronger hypotheses. And yet this effect is often sufficient to induce a qualitative change in the equilibrium outcome -- namely, to stabilize good science in the face of pressure to publish.

Like all models, ours is a simplification and abstraction of what is, in reality, an incredibly complex process. The purpose of the model is to cut through the complexity whilst retaining the most salient forces at play when scientists make decisions about what to study and by what methodology and effort. The value of a mathematical or computational model over a verbal hypothesis is that it allows systematic exploration of how these fundamental forces play out, without relying on intuition alone \citep{smaldino2017models}.

To be truly useful, a model should tell us something that we did not know before we built it
-- that is, the model need not be entirely stupid \citep{smaldino2017models}. 
In the context of scientific practice, we have seen that theory must provide new information about what constitutes a strong versus a weak hypothesis, in order to promote good science. A theoretical model whose output simply recapitulates the assumptions that went into building it is tautological, and it does not grant us any additional ability to distinguish between strong and weak hypotheses; it is wasted effort that does not help promote good science.

Our findings reinforce and justify the calls made by several authors for better theory, particularly in the social sciences \citep{smaldino2016natural,muthukrishna2019problem, smaldino2019better}. We also offer some optimistic results for those who lament the pressure to publish as corroding good science \citep{Sarewitz:2016aa,Rawat2014,Oliveira2015,Serhat2018}. Such concerns have a long history \citep{Price1963-PRILSB-3} and an exponentially expanding scientific literature \citep{Bornmann} poses profound challenges for researchers, even if the rate of false-positive reports is low. Yet our results show that pressure to publish, and competition between labs in general, can stimulate effort and produce excellent science provided the theoretical and empirical tools in a field are sufficiently well developed. It is only when theoretical tools are not yet developed, or go unused, that publication pressures create perverse incentives and lead to the evolution of bad science. 

\clearpage

\setcounter{equation}{0}

\begin{centering}
{\huge
\textbf{Supplementary information}
}
\bigskip
\\
\bigskip
\end{centering}

\tableofcontents
\newpage

\noindent In this supplement we provide derivations for the equations in the main text, along with additional simulation results to demonstrate the robustness of our findings to relaxation of model assumptions. 

\section{Adaptive dynamics model of publication}

We analyze a model for the natural selection of scientific publication strategy under the framework of adaptive dynamics \cite{Mullon:2016aa,Leimar}. Within this framework we follow the basic assumptions of \cite{smaldino2016natural}: a lab's success is measured in terms of the number of publications and (un)successful replications of their work by other labs. We assume that labs  ``reproduce'' by adopting the research strategies of other labs, based on their past success. Under this framework we assume an infinite population of labs, each using the same resident publication strategy, and we perform an invasion analysis to determine which resident strategies are stable in the face of local ``mutations'' that perturb the resident research strategy. While the assumptions of adaptive dynamics are unrealistic in several important ways, they nonetheless allow us to systematically explore the qualitative behavior of the system, and our key finding from the analysis -- that competition to publish can produce good science when the role of theory in selecting hypotheses is accounted for -- holds  when relaxation these simplifying assumptions in individual-based simulations.

\subsection{Lab life cycle}

We consider a population of labs whose life cycle proceeds via a phase of publication followed by a phase of selection and reproduction, in which the current population is replaced with a population of new labs. This simplifying assumption allows us to assume that all labs are the same age during the selection phase and ignore effects that arise due to older labs appearing more successful due to having had more time to publish. We relax this assumption in our simulations and show that it does not qualitatively alter our results.

As described in the main text, a lab $i$ produces novel results at a rate 

\begin{equation}
\rho_i=(1-\eta\log_{10}(e_i))\times(1-r_i)\times(P_i(T)P_i(+|T)+P_i(F)P_i(+|F)),
\end{equation} 
\\
The probability that the hypothesis being tested is true is given by 

\begin{equation}
P_i(T)=\frac{b_0+b_1(e_i-1)}{e_i}
\end{equation}
\\
where $P_i(F)=1-P_i(T)$ and

\begin{align}
\nonumber &P_i(+|T)=\frac{V_i}{\gamma}\times\frac{\gamma e_i}{1+\gamma(e_i-1)}\\
&P_i(+|F)=\frac{V_i}{\theta}\times\frac{1+(\theta-1)e_i}{1+(\theta-V)(e_i-1)}.
\end{align}
\\
Labs produce replication studies at rate

\begin{equation}
\phi_i=(1-\eta\log_{10}(e_i))\times r_i
\end{equation}
\\
and they reproduce the original finding of a study by lab $j$ with probability

\begin{equation}
p_{ij}=\frac{P_i(T)P_i(+|T)P_j(+|T)+P_i(F)P_i(+|F)P_j(+|F)}{P_i(T)P_i(+|T)+P_i(F)P_i(+|F)}.
\end{equation}
\\
while they produce a different finding to lab $j$ with probability

\begin{equation}
q_{ij}=\frac{P_i(T)P_i(+|T)(1-P_j(+|T))+P_i(F)P_i(+|F)(1-P_j(+|F))}{P_i(T)P_i(+|T)+P_i(F)P_i(+|F)}.
\end{equation}
\\
We also define $p_i=\frac{1}{N-1}\sum_{j\neq i}p_{ij}$ and $q_i=\frac{1}{N-1}\sum_{j\neq i}q_{ij}$, the probability of successful and unsuccessful replication attempts for lab $i$ by the rest of the population.
We can now model the publication phase as a system of ODEs with $x^i_n(t)$ the number of novel results that have been produced at time $t$ by lab $i$ and $x_r^i$ the number or replication studies published by lab $i$. We also define $z^i(t)$ as the number of novel studies produced by lab $i$ that have been replicated by other labs at time $t$. Under these assumptions the dynamics of publication are as follows

\begin{eqnarray}
\nonumber\frac{d x^i_n}{dt}&=&\rho_i\\
\nonumber\frac{d x^i_r}{dt}&=&\phi_i\\
\frac{d z^i}{dt}&=&\sum_{j\neq i}\frac{x^i_n-z^i}{L}\phi_j
\end{eqnarray}
\\
where $L$ is the size of the corpus of published materials available for replication which is assumed for simplicity to be fixed. Assuming a monomorphic population such that $\phi_j=\phi$, and setting the number of publications at time $t=0$ to zero, the distribution of publications for a lab $i$ at time $t$ is given by

\begin{eqnarray}
\nonumber x^i_n(t)&=&\rho_it\\
\nonumber x^i_r(t)&=&\phi_it\\
z^i(t)&=&(N-1)\left(\mathrm{e}^{-\phi t/L}-1+\frac{\phi}{L}t\right)\frac{L\rho_i}{\phi}
\end{eqnarray}
\\
We assume that the lifespan of each lab is one time unit, such that the integral must be evaluated at $t=1$. This corresponds to a  scenario in which there are many more publications in the corpus of literature for a field than can be replicated in the lifetime of a lab. By Taylor expanding $z^i(t)$ in terms of $L^{-1}$ and neglecting terms $O\left(L^{-2}\right)$ and higher we recover

\begin{eqnarray}
\nonumber x^i_n&=&\rho_i\\
\nonumber x^i_r&=&\phi_i\\
z^i&=&(N-1)\frac{1}{2}\frac{\phi\rho_i}{L}+O\left(L^{-2}\right)
\end{eqnarray}
\\
and taking the limit $N\to\infty$, $L\to\infty$ and $L/N\to l$ we recover 

\begin{eqnarray}
\nonumber x^i_n&=&\rho_i\\
\nonumber x^i_r&=&\phi_i\\
z^i&=&\frac{1}{2}\frac{\phi\rho_i}{l}
\end{eqnarray}
\\

\subsection{Invasion analysis}

Taking Eq.~10 as the publication distribution at the end of the publication cycle, we can then describe the fitness of a lab $i$ against a monomorphic background of competing labs, following publication as

\begin{equation}
w(e_i,V_i,r_i)=\rho_iB_N+\phi_iB_r+
\frac{1}{2}\frac{\rho_i\phi}{l}(p_{i}B_{O+}-q_{i}C_{O-})
\end{equation}
\\
which we can write as

\begin{eqnarray}
\nonumber w(e_i,V_i,r_i)=\\
\nonumber (1-\eta\log_{10}(e_i))\times(1-r_i)\Bigg[P_i(T)P_i(+|T)\left(B_N+\frac{\phi}{2l}B_{O+}P(+|T)-\frac{\phi}{2l}C_{O-}(1-P(+|T))\right)+\\
\nonumber P_i(F)P_i(+|F)\left(B_N+\frac{\phi}{2l}B_{O+}P(+|F)-\frac{\phi}{2l}C_{O-}(1-P(+|F))\right)\Bigg]+(1-\eta\log_{10}(e_i))\times r_i B_r\\
\end{eqnarray}
\\
We can now compute the selection gradient for the system. From Eq.~3 we immediately see that fitness is monotonically increasing in $V_i$ thus we need only evaluate the gradient at $w(e_i,1,r_i)$. This gives us

\begin{eqnarray}
\nonumber s_e=\frac{\partial w}{\partial e_i}\Bigg |_{e_i=e,r_i=r}=&-\frac{\eta}{e\log[10]}\Bigg[(1-r)P(T)P(+|T)\alpha+ (1-r)P(F)P(+|F)\beta+rB_r\Bigg]+\\
&(1-\eta\log_{10}(e))\times(1-r)\Bigg[\frac{d (P_i(T)P_i(+|T))}{de_i}\alpha+\frac{d (P_i(F)P_i(+|F))}{de_i}\beta\Bigg]\\
\nonumber s_r=\frac{\partial w}{\partial r_i}\Bigg |_{e_i=e,r_i=r}=&-(1-\eta\log_{10}(e))\Bigg[P(T)P(+|T)\alpha+ P(F)P(+|F)\beta\Bigg]+(1-\eta\log_{10}(e)) B_r\\
\end{eqnarray}
\\
where

\begin{eqnarray}
\nonumber\alpha&=&\left(B_N+\frac{\phi}{2l}B_{O+}P(+|T)-\frac{\phi}{2l}C_{O-}(1-P(+|T))\right)\\
\beta&=&\left(B_N+\frac{\phi}{2l}B_{O+}P(+|F)-\frac{\phi}{2l}C_{O-}(1-P(+|F))\right)
\end{eqnarray}
\\
with

\begin{eqnarray}
\frac{d (P_i(T)P_i(+|T))}{de_i}=\frac{b_1-\gamma b_0}{(1+\gamma(e_i-1))^2}
\end{eqnarray}
\\
and

\begin{eqnarray}
\nonumber\frac{d (P_i(F)P_i(+|F))}{de_i}=-\frac{1}{1+(\theta-1)(e_i-1)}\frac{1}{\theta e_i}\times\\
\nonumber\left[(1+(\theta-1)e_i)\frac{(b_1-b_0)}{e_i}+(e_i(1-b_1)+(b_1-b_0))\left(\frac{\theta-1}{1+(\theta-1)(e_i-1)}\right)\right].\\
\end{eqnarray}
\\
From Eqs.~13-17 we can calculate the points at which the selection gradient vanishes, $(\hat{e},\hat{r})$, which satisfy:

\begin{eqnarray}
\nonumber\frac{\eta}{\hat{e}\log[10]}\Bigg[\left(\frac{b_0+b_1(\hat{e}-1)}{\hat{e}}\right)\left(\frac{\hat{e}}{1+\gamma(\hat{e}-1)}\right)\alpha+\\
\nonumber\left(1-\frac{b_0+b_1(\hat{e}-1)}{\hat{e}}\right)\left(\frac{1+(\theta-1)\hat{e}}{\theta(1+(\theta-V)(\hat{e}-1))}\right)\beta+\frac{\hat{r}}{1-\hat{r}}B_r\Bigg]=\\
\nonumber(1-\eta\log_{10}(\hat{e}))\times\Bigg[\left(\frac{b_1-\gamma b_0}{(1+\gamma(\hat{e}-1))^2}\right)\alpha-\frac{1}{1+(\theta-1)(\hat{e}-1)}\frac{1}{\theta \hat{e}}\times\\
\nonumber\left[(1+(\theta-1)\hat{e})\frac{(b_1-b_0)}{\hat{e}}+(\hat{e}(1-b_1)+(b_1-b_0))\left(\frac{\theta-1}{1+(\theta-1)(\hat{e}-1)}\right)\right]\beta\Bigg]\\
\end{eqnarray}
\\
where

\begin{eqnarray}
\nonumber \hat{r}=&\\
\nonumber &\frac{B_r-B_N(\hat{P}(T)\hat{P}(+|T)-\hat{P}(F)\hat{P}(+|F))}{(B_{O+}\hat{P}(+|T)-C_{O-}(1-\hat{P}(+|T)))\hat{P}(T)\hat{P}(+|T)+(B_{O+}\hat{P}(+|F)-C_{O-}(1-\hat{P}(+|F)))\hat{P}(F)\hat{P}(+|F)}\times\frac{2l}{(1-\eta\log_{10}(\hat{e}))}.\\
\end{eqnarray}
\\
Eqs.~18-19 cannot be solved analytically in general and in particular Eq.~18 can produce multiple solutions in the physically relevant range. However we observe that the condition for any equilibrium to be convergent stable under all mutation matrices is that the $2\times2$ Jacobian matrix $\mathbf{J}$ for the system must have negative eigenvalues or, equivalently, be negative definite \citep{Leimar} which in turn implies that $(\mathbf{J})_{rr}=\frac{\partial s_r}{\partial r}<0$ must hold. This condition is satisfied only if
\begin{eqnarray}
\nonumber P(T)P(+|T)(B_{O+}P(+|T)-C_{O-}(1-P(+|T)))+\\
P(F)P(+|F)(B_{O+}P(+|F)-C_{O-}(1-P(+|F)))>0.
\end{eqnarray}
\\
If we assume $C_{O-}\gg B_{O+}$, i.e.~the penalties for publishing false results are very large, then Eq.~20 is only satisfied in the limit $P(+|T)\to 1$ and $P(+|F)\to 1$ which is the bad-science equilibrium. Thus under our model assumptions there are no points of zero selection gradient that are convergent stable except close to the bad-science equilibrium. This is consistent with our numerical analysis of the system (Figure S1), under which we find only unstable singular points. However this result does not exclude the possibility that stable equilibria can arise at the boundaries of phase space.

\subsection{Boundary behavior}

Stable equilibria can arise at the boundary if the selection gradient perpendicular to the boundary points towards it, and the selection gradient parallel to the boundary is zero. We now explore the behavior of the system at the boundaries $r=0$, $r=1$, $(e-1)/e=0$ and $(e-1)/e=1$ beginning with tje resident bad-science equilibrium of \cite{smaldino2016natural} which corresponds to $(e-1)/e=r=0$.
\\
\\
\textbf{Bad science $(e=1,r=0)$:}
The bad-science equilibrium of \cite{smaldino2016natural} arises at $(e=1,r=0)$. From Eqs.~13-14 the selection gradient at this point is

\begin{eqnarray}
\nonumber s_e(1,0)&=&-\frac{\eta}{e\log[10]}B_N-
(1-\eta\log_{10}(e))\times\Bigg[b_0(\gamma -1)+(1-b_0)(\theta-1)^2/\theta\Bigg]B_N\\
\nonumber s_r(1,0)&=&-(1-\eta\log_{10}(e))(B_N-B_r)\\
\end{eqnarray}
\\
which is always negative, indicating that the bad-science equilibrium is always a stable state of the system provided the benefit of publishing a novel result is greater than that for publishing a replication, $B_N\geq B_r$.
\\
\\
\textbf{Maximum replication $(r=1)$:} When replication rate is at its maximum, $r=1$, the selection gradient parallel to the boundary, calculated from Eq.~13, is given by

\begin{eqnarray}
s_e(e,1)=-\frac{\eta}{e\log[10]}B_r
\end{eqnarray}
\\
which is always negative. Thus we need only evaluate the selection gradient perpendicular to the boundary at $(r=1, e=1)$ which, from Eq.~14 gives

\begin{eqnarray}
\nonumber s_r(1,1)&=&-\Bigg[B_N +B_{O+}/l-B_r\Bigg]
\end{eqnarray}
\\
which, under our assumption $B_N\geq B_r$, is always negative. Thus there is no stable equilibrium with maximum replication.
\\
\\
\textbf{Minimum replication $(r=0)$:} Finally we consider the behavior of the system when replication rate is minimized, $r=0$. For Eqs.~13-14 we find selection gradient

\begin{eqnarray}
\nonumber &s_e(e,0)=-\frac{\eta}{e\log[10]}\Bigg[(P(T)P(+|T)+ P(F)P(+|F)\Bigg]B_N+\\
&(1-\eta\log_{10}(e))\times\Bigg[\frac{d (P_i(T)P_i(+|T))}{de_i}+\frac{d (P_i(F)P_i(+|F))}{de_i}\Bigg]B_N\\
\nonumber &s_r(e,0)=-(1-\eta\log_{10}(e))\Bigg[(P(T)P(+|T)+ P(F)P(+|F))B_N-B_r\Bigg]\\
\end{eqnarray}
\\
at the boundary, where Eq.~23 must be solved numerically as above. Eq.~24 is negative provided $(P(T)P(+|T)+ P(F)P(+|F))B_N>B_r$. From Eq.~16-17, $(P(T)P(+|T)+ P(F)P(+|F))$ is non-monotonic in $e$ and thus, depending on the solution to Eq.~23 and the choice of $B_r$ the equilibrium may be either stable or unstable. Crucially this means that the addition of replication to the evolutionary dynamics of the system may cause a stable, high-effort equilibrium to become unstable.

The resulting evolutionary trajectories of the system across a range of parameter are shown in Figure S1 and the basin of attraction for the good- and bad-science equilibria in the absence of replication are shown for different model parameters in Figure S2.

\begin{figure}[th!] \centering \includegraphics[scale=0.2]{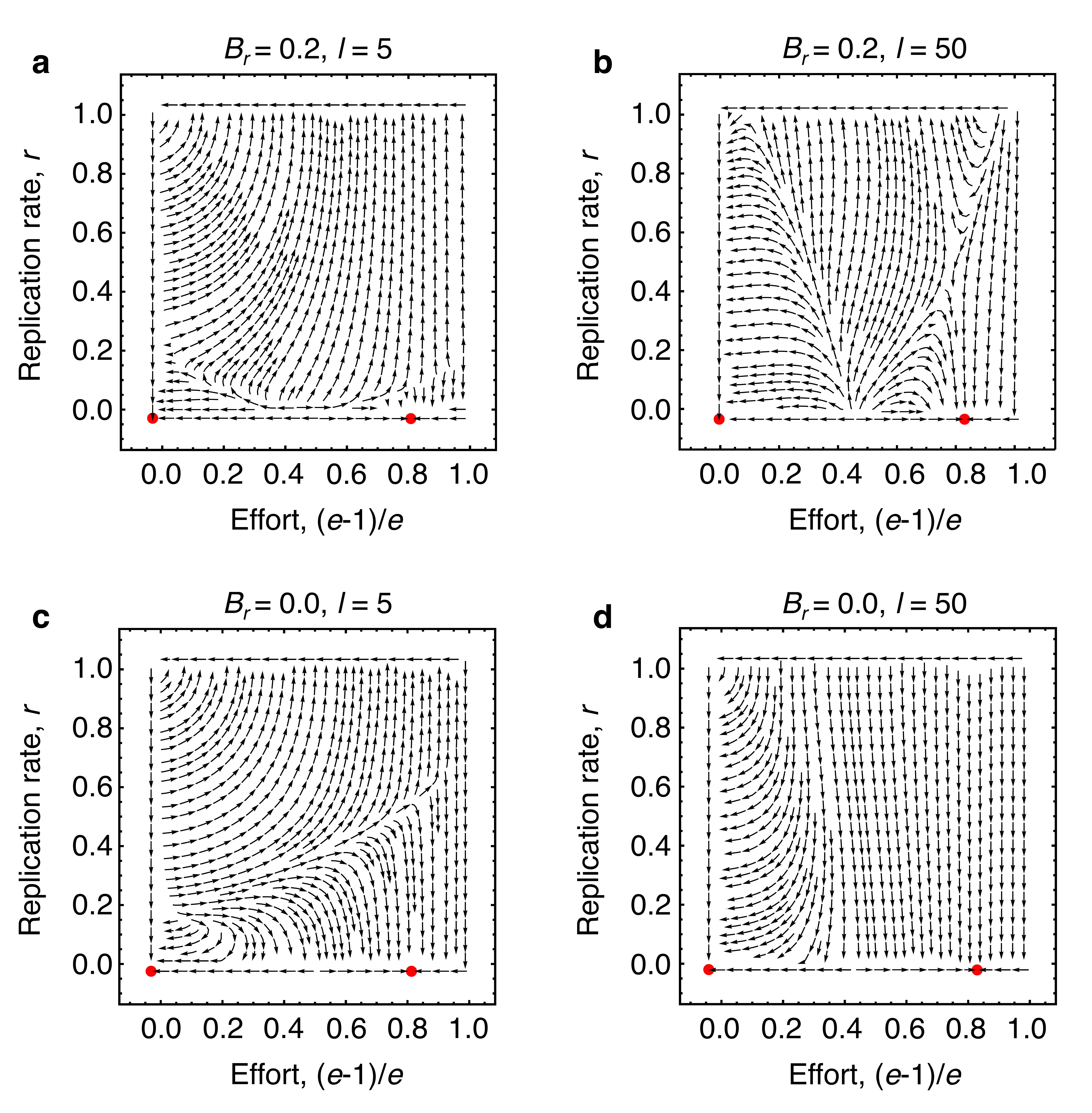}
\caption*{\small \textbf{Figure S1: Co-evolution of $e$ and $r$} Phase portraits in the regime of adaptive dynamics for a) high benefits for replication, $B_r=0.2$ and a small corpus of literature $l=5$ b) high benefits for replication, $B_r=0.2$ and a large corpus of literature $l=50$ c) no benefit for replication, $B_r=0.0$ and a small corpus of literature $l=5$ b) no benefit for replication, $B_r=0.0$ and a large corpus of literature $l=50$. All other parameters are chosen as in Figure 3. The  good-science equilibrium consisting of high effort and zero replication rates, always exists alongside the bad-science equilibrium at minimum effort and zero replication rate. We see that high levels of replication can undermine good science and pull the system back to the bad-science equilibrium . Both equilibria are marked with red dots. In all cases we assume that the costs for failed replication is high, $C_{O-}=100$ }
\end{figure}

\begin{figure}[th!] \centering \includegraphics[scale=0.2]{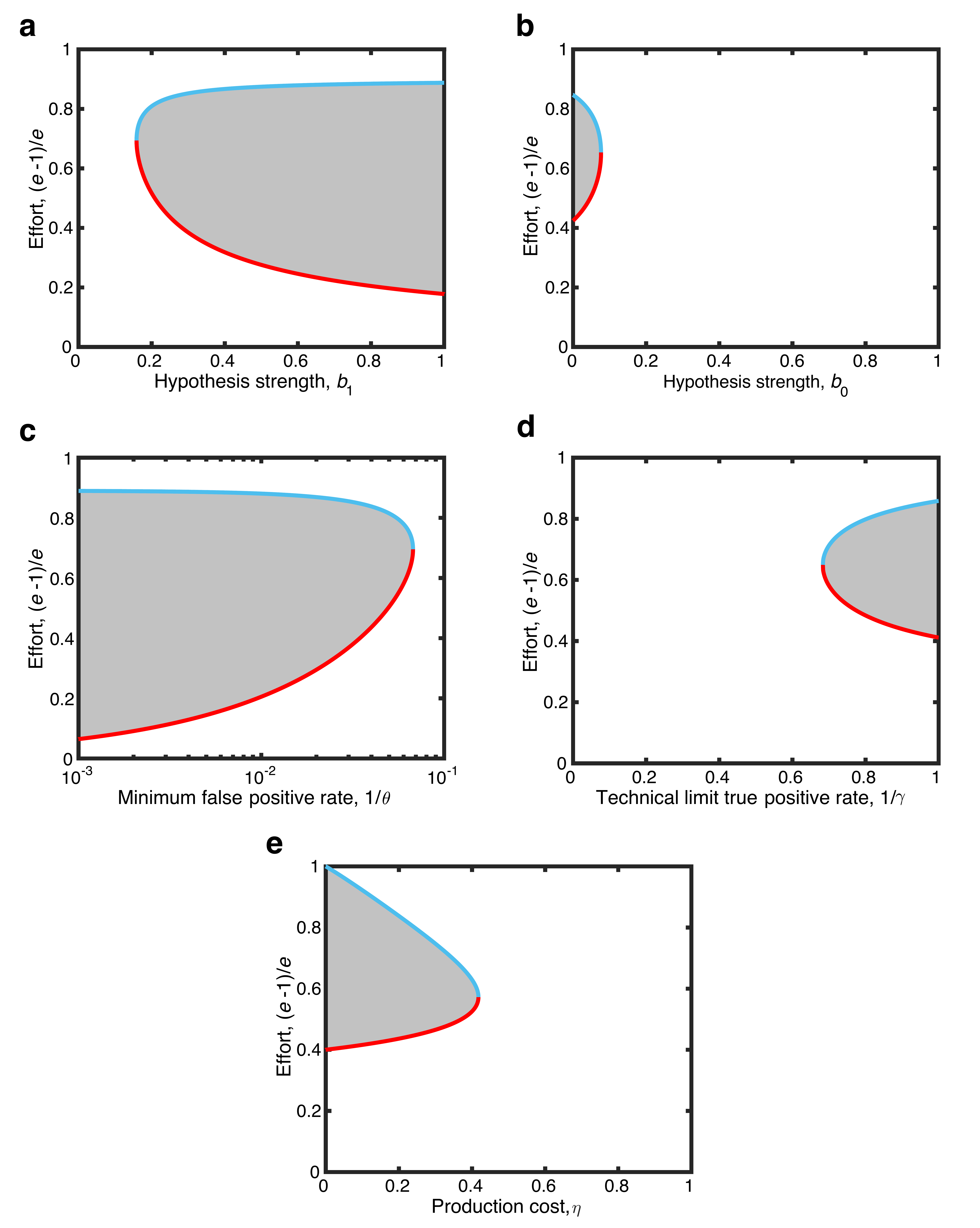}
\caption*{\small \textbf{Figure S2: Analysis of equillibria by adaptive dynamics.} The figure shows equilibrium publication strategies in a large population of labs, as a function of model parameters. Plotted in each panel are the locations of the stable (blue) and unstable (red) equilibria as a function of all five parameters of the system without replication. For many parameter choices the system is bi-stable, with a good-science equilibrium indicated by the blue line and a bad-science equilibrium at minimum effort $(e-1)/e=0$. In the gray regions selection favors increasing effort towards the good-science equilibrium; whereas in the white regions selection favors ever decreasing effort towards to bad-science equilibrium. a) Impact of hypothesis strength $b_1$ of the basin of attraction for good science. b) Impact of hypothesis strength $b_0$. c) Impact of the technical limit false-positive rate $1/theta$. d) Impact of the technical limit true-positive rate $1/\gamma$. e) Impact of the production cost of science $\eta$. Payoffs are set at $B_N=1$, and $l=1$. All other parameters are as in Figure 2 unless otherwise specified in the panel.}
\end{figure}

\clearpage

\subsection{Replication as a policy}

So far we have studied replication as an evolving trait, which labs can choose to engage in as a way to improve their success through publication. However replication of published research can, in principle at least, be implemented as policy, in which a proportion $r$ of all published studies are replicated by an outside agency. To study replication as policy it is sufficient to set $B_r=0$ and $r_i=0$ in Eq.~12. We then retrieve selection gradient 

\begin{eqnarray}
\nonumber s_e=\frac{\partial w}{\partial e_i}\Bigg |_{e_i=e}=&-\frac{\eta}{e\log[10]}\Bigg[P(T)P(+|T)\alpha+ P(F)P(+|F)\beta\Bigg]+\\
\nonumber&(1-\eta\log_{10}(e))\times\Bigg[\frac{d (P_i(T)P_i(+|T))}{de_i}\alpha+\frac{d (P_i(F)P_i(+|F))}{de_i}\beta\Bigg]\\
\end{eqnarray}
\\
where $\alpha$ and $\beta$ given by Eq.~15 account for the amount of enforced replication under the policy. As in previous examples, Eq.~25 must be solved numerically. Figure S3 shows the basin of attraction for good and bad science as a function of replication rate $r$ and literature size $l$. We see that more stringent replication (arising from either higher rates of enforced replication, or a lower ratio of literature to labs) results in a larger basin of attraction for good science.

\begin{figure}[th!] \centering \includegraphics[scale=0.15]{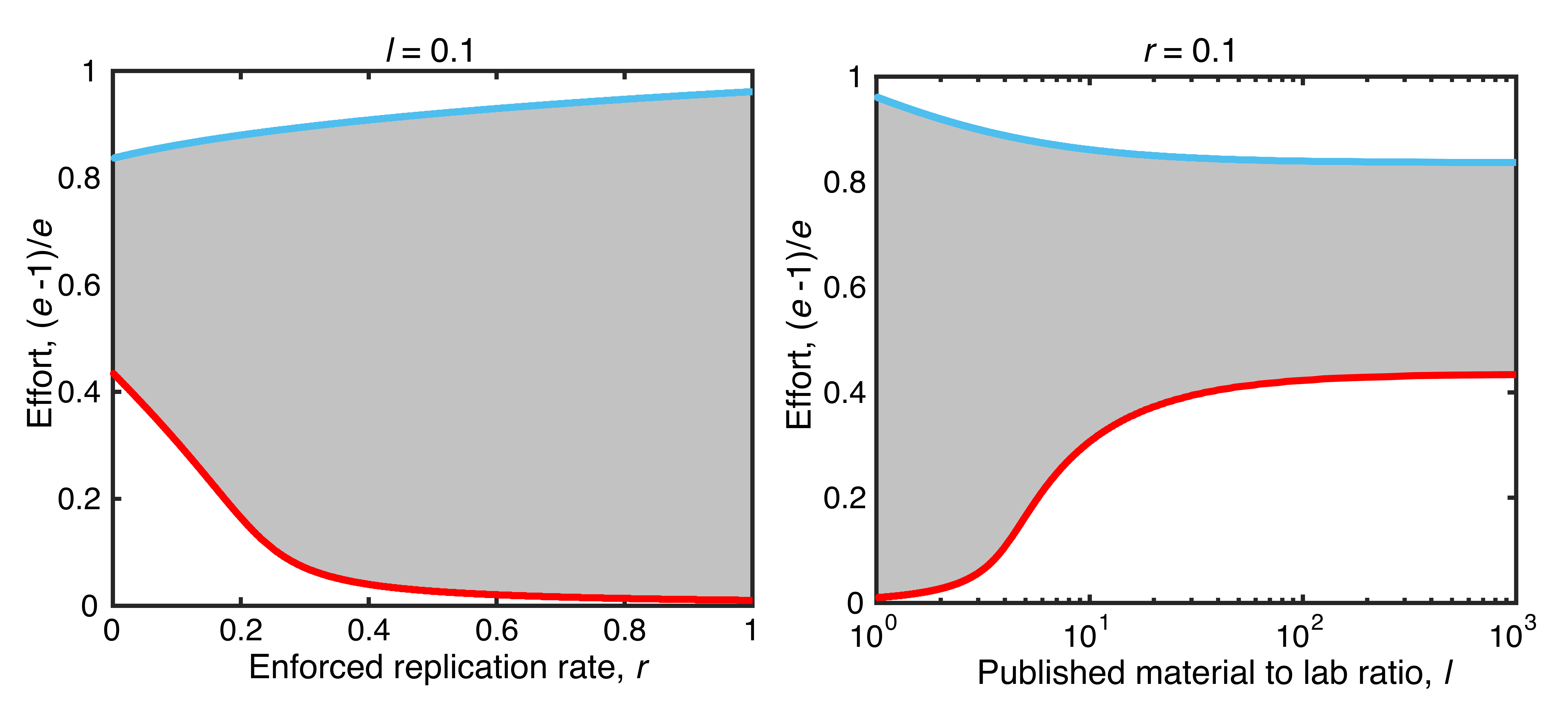}
\caption*{\small \textbf{Figure S3: Replication as a policy under adaptive dynamics.} The figure shows the same information as in Figure S2 under a scenario where replication occurs at a fixed rate $r$ and $B_r=0$. a) Impact of replication rate on the basin of attraction for good science for a corpus size $l=5$. b) Impact of corpus size $l$ on the basin of attraction for good science for a fixed replication rate $r=0.1$. All other parameters are as in Figure 2.}
\end{figure}

\clearpage

\section{Individual-based simulations}

We ran individual-based simulations, relaxing the assumptions of the adaptive dynamics model described above to account for (i) variation in lab age and (ii) heterogeneity in lab publication strategy (iii) a finite population of active labs. We treated effort $e$, efficacy $V$ and replication rate $r$ as heritable, evolving traits. We ran ensembles of $10^3$ replicate simulations to produce each simulation figure and plotted the average trajectories over time. Further details of the simulation setup are provided below.

\subsection{Lab aging}
Under the assumptions of adaptive dynamics the population of labs is infinite and the lab life cycle ensures that all labs are the same age when natural selection occurs. These simplifying assumptions are made for mathematical convenience but do not describe a particularly realistic case: in any given field there is a wide range of labs of different ages, and the older a lab is, the more it has published. This has consequences for the rate at which the lab experiences replication attempts (as they have contributed more novel results to the corpus of results in their field) which in turn has consequences for their fitness.

We assume that labs ``die'' when they copy another lab's strategy (see below). Furthermore we assume that the fitness of a lab is determined by the average payoff received due to novel publication and replication over the lab lifetime.

\subsection{Natural selection and the copying process}

We assume that lab birth and death occurs via the copying process \cite{Traulsen:2006zr} used to study a process of cultural evolution via imitation. Under this model, we assume that a pair of labs $i$ and $j$ are chosen at random, such that lab $i$ chooses to adopt the strategy of lab $j$ with a probability $\pi_{ij}$ where

\begin{equation}
\pi_{ij}=\frac{1}{1+e^{\sigma(\bar{w}_i-\bar{w}j)}}
\end{equation}
\\
where $\bar{w}_i$ is the average payoff to lab $j$ during its lifetime. This birth-death process can be thought of as a fixed population of labs who update their strategies, described by their methodological efficacy $V$, effort $e$ and replication rate $r$, when they see another lab doing better. This may be thought of as occurring whenever an old lab is disbanded and replaced with a new lab in a university or research institute. Alternatively it may be understood as occurring among a fixed population of competing labs trying to gain an edge over one another. 

\subsection{Replication}

Populations of competing labs are assumed to contribute to a corpus of literature of size $L$. When choosing a study to replicate a lab chooses a study at random from the corpus. They attempt to reproduce the study using the same level of methodological efficacy $V$ and effort $e$ as for testing a novel hypothesis. After an attempt at reproduction the study is moved from the corpus of literature available for replication.

As a result, a lab that has produced $n$ papers with novel results has a study reproduced with probability $n/L$ when another lab decides to undertake a replication study. If the outcome of the replicating labs study is positive, the replication is successful otherwise it is not. The corpus of literature is always assumed to contain $L$ novel papers available for replication - if all the papers by currently active labs have been replicated we assume that the labs can still reproduce older literature. Thus labs can in principle engage in replication at the maximum rate $r=1$, although this pathological case is not observed in simulations or in under the adaptive dynamics model, except transiently (Figure S1).

\subsection{Co-evolution of effort and replication}

We explored the co-evolutionary dynamics of replication and effort via individual-based simulations (Figure S4). In the absence of hypothesis choice only very low levels of replication emerged and, as in Figure 2 and Figure 4 of the main text, effort evolved to the bad-science minimum. In contrast, when hypothesis choice was allowed the good-science equilibrium was maintained and replication evolved steadily to around $r=0.1$ (Figure S4b).

\begin{figure}[th!] \centering \includegraphics[scale=0.2]{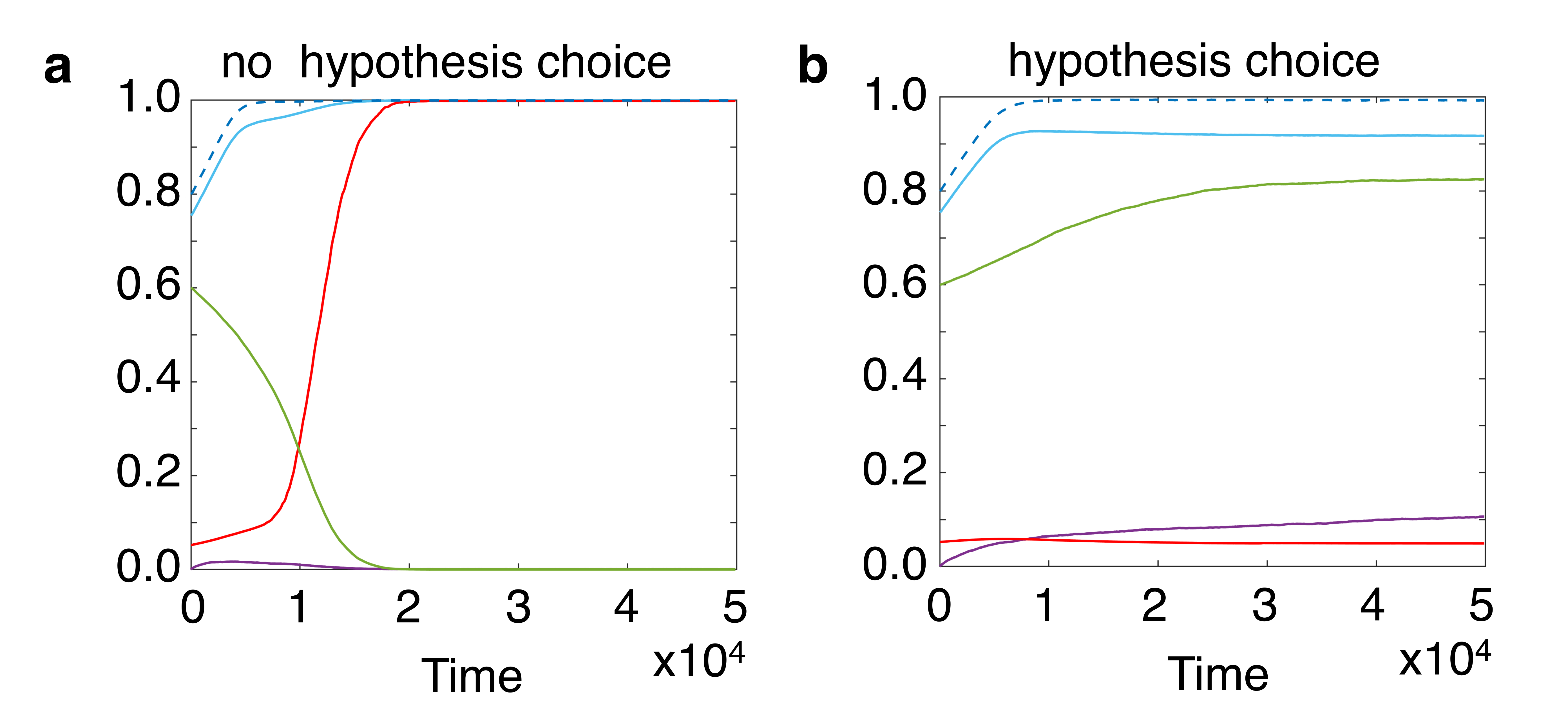}
\caption*{\small \textbf{Figure S4: Co-evolution of replication and effort.} The figure shows results of individual-based simulations for the co-evolution replication and effort. (a) In the absence of hypothesis choice, both true (blue) and false (red) positive rates increase to unity, and effort declines to a minimum $(e-1)/e=0$ (green), while replication rate (purple) remains low. b) However, when hypothesis choice is allowed, effort increases over time towards a good-science equilibrium in which false positives are rare, and replication evolves to a modest rate. All parameters are the same as in Figure 2c. Replications are chosen from a corpus of $L=10^5$ novel studies, and each study is allowed to be replicated only once (see SI). Payoffs are $B_N=1$, $B_r=0.2$, $B_{O+}=0.1$ and $C_{O-}=100$.}
\end{figure}

\subsection{Limit of $\theta=\gamma=1$}

Our model reproduces that of \cite{smaldino2016natural} in the limit $\gamma=\theta=1$, and as such our simulations in this limit should produce the same qualitative results. We ran simulations in this limit without hypothesis choice and showed that, indeed, the bad-science equilibrium quickly emerged (Figure S5a). When hypothesis choice was allowed (Figure S5b) the bad-science equilibrium still evolved in this limit, since power $P(+|T)$ and false positive rate $P(+|F)$ are \textit{both} independent of effort under this model, once efficacy evolves to its maximum $V=1$. This latter result illustrates a pathology of the limit $\theta=\gamma=1$, under which bad science (true- and false-positive rates equal to one) cannot be avoided, no matter how much effort a lab puts in, once methodological efficacy reaches its maximum -- a state of affairs that does not reflect reality in any scientific field. However, when we separate out methodological efficacy from lab effort in identifying positive results, and allow for the possibility that a diligent lab can, in principle, expend effort to do good science (i.e.~by setting $\gamma>1$ and $\theta>1$), the effects of theory on stabilizing good science become apparent, and both good- and bad-science equilibria emerge -- a state of affairs that more accurately reflects what we see in scientific practice across fields. 

\begin{figure}[th!] \centering \includegraphics[scale=0.2]{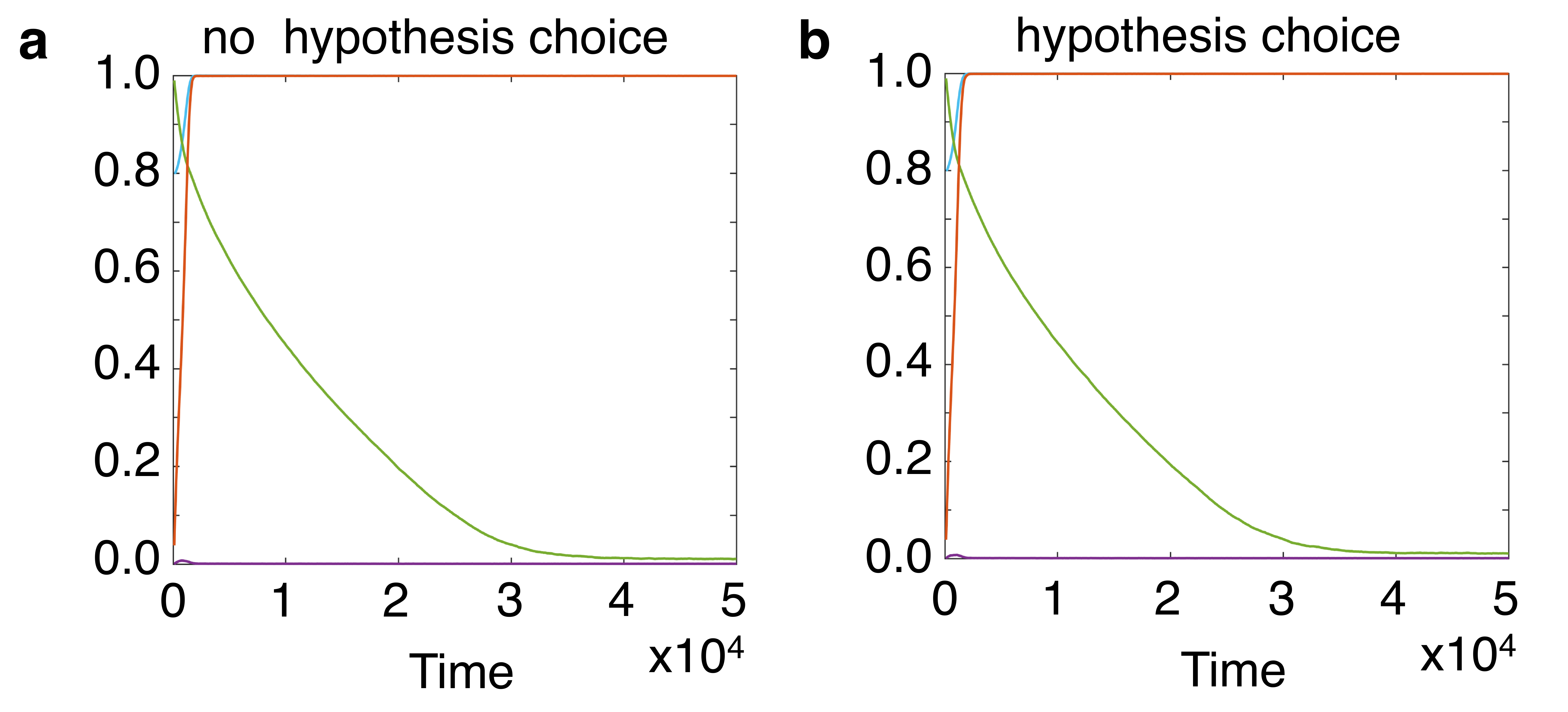}
\caption*{\small \textbf{Figure S5: Simulations in the limit $\theta=\gamma=1$.} This figure is the same as Figure S4 with the alteration that the technical limits of false- and true-positives are set to $\theta=\gamma=1$. In this case both without (a) and with (b) hypothesis choice, bad science evolves.}
\end{figure}

\end{document}